\documentclass[%
 aip,
 amsmath,amssymb,
reprint,%
]{revtex4-1}
\usepackage{graphicx}
\usepackage{dcolumn}
\usepackage{bm}
\usepackage{color}

\usepackage[utf8]{inputenc}
\usepackage[T1]{fontenc}
\usepackage{mathptmx}
\usepackage{subfig}
\usepackage{amsmath}
\usepackage[inline]{enumitem}
\usepackage{caption}
\usepackage{mathrsfs}
\usepackage{comment}

\usepackage{float}
\newcommand{\cross}{\times}

\usepackage[normalem]{ulem}

\begin{document}

\title{Similarity for downscaled kinetic simulations of electrostatic plasmas: reconciling the large system size with small Debye length}

\author{Yanzeng Zhang}
\affiliation{Los Alamos National Laboratory, Los Alamos, NM 87545, USA}

\author{Haotian Mao}
\affiliation{University of California at San Diego, La Jolla, California 92093, USA
}
\affiliation{Los Alamos National Laboratory, Los Alamos, NM 87545, USA}

\author{Yuzhi Li}
\affiliation{Los Alamos National Laboratory, Los Alamos, NM 87545, USA}

\author{Xian-Zhu Tang}
\affiliation{Los Alamos National Laboratory, Los Alamos, NM 87545, USA}

\begin{abstract}
A simple similarity has been proposed for kinetic (e.g.,
particle-in-cell) simulations of plasma transport that can effectively
address the longstanding challenge of reconciling the tiny Debye
length with the vast system size.  This applies to both transport in
unmagnetized plasma and parallel transport in magnetized plasmas,
where the characteristics length scales are given by the Debye length,
collisional mean free paths, and the system or gradient lengths. The controlled scaled variables are the configuration space, $\mathbf{x}/\mathscr{L},$ and \textcolor{black}{artificial collisional rates, $\mathscr{L}\mu$, which is realized through scaling the Coulomb Logarithm in the simulations}, $\mathscr{L}\ln \Lambda.$ Whereas, the scaled time,
$t/\mathscr{L}$, and electric field, $\mathscr{L}\mathbf{E}$, are automatic outcomes.  The similarity properties are examined, demonstrating that the
macroscopic transport physics is preserved through a similarity transformation while
keeping the microscopic physics at its original scale of Debye length. To showcase the utility
of this approach, two examples of 1D plasma transport problems were
simulated using the VPIC code: the plasma thermal quench in tokamaks
[J. Li, et al., Nuclear Fusion \textbf{63}, 066030 (2023)] and the
plasma sheath in the high-recycling regime [Y. Li, et al., Physics of
  Plasmas \textbf{30}, 063505 (2023)].
\end{abstract}

\maketitle

First-principles kinetic plasma simulations, with either the
Particle-In-Cell
(PIC)~\cite{birdsall2018plasma,tajima2018computational,verboncoeur2005particle}
or the
continuum~\cite{cheng1976integration,filbet2003comparison,cheng2014discontinuous}
approach, can provide critical insights into complex plasma dynamics,
and have served as a cornerstone of high-fidelity plasma
modelings. They are particularly in demand for nearly collisionless
plasmas, where the plasma mean free path, $\lambda_{mfp} \propto
T_e^2/n_e$, is much longer than the characteristic system size, $L$,
as a result of the high plasma temperatures ($T_e$), low plasma
densities ($n_e$), or a combination of the
two~\cite{ITER-transport-nf-2007,Fernandez-etal-nf-2022,NAP-2010,Parks-ency-atom-science-2015,Zhuravleva-etal-nature-2014}. Even
in the opposite limit of collisional plasmas with $\lambda_{mfp}\ll
L$, kinetic effects can still play an important role, especially when
the electron distribution function deviates significantly from a local
Maxwellian. These include the plasmas in the sheath and presheath
regions, where the electron distribution has a one-sided cutoff in
parallel velocity with respect to the magnetic field due to the
electrostatic trapping effect of the sheath/presheath electric field
that reflects thermal electrons for
ambipolarity~\cite{guo2012ambipolar,tang2016kinetic}. An even more
dramatic situation of spatially-extended non-Maxwellian distribution
functions arises when a nearly collisionless plasma interacts with a
cold, dense, and often collisional plasma, e.g., in the plasma thermal quench (TQ) of a major disruption in
tokamaks~\cite{zhang2023cooling,zhang2023electron,li2023staged} and
the formation of structures in a galaxy
clusters~\cite{Fabian-ARAA-1994,Peterson-Fabian-PR-2006}.

It is often the case that these high-fidelity kinetic simulations can
incur extreme computational costs. The difficulty comes about because
the kinetic simulation has to reconcile a large system size, $L$, with
the small plasma Debye length, $\lambda_D$. This would be a physics
fidelity requirement in physical systems where a localized non-neutral
plasma structure is present in an otherwise quasi-neutral plasma. Such
non-neutral plasma structures would have a width on the order of the
Debye length due to the plasma shielding
effect~\cite{chen2012introduction}. Well-known examples include the
sheath in a plasma intercepting a
solid~\cite{Langmuir,langmuir-pra-1929,lieberman-lichtenberg-book-2005,stangeby-book-2000}
and the parallel collisionless
shocks~\cite{zhang2023cooling,taylor1970observation,romagnani2008observation,forslund1970formation,sarri2010shock,moiseev1963collisionless},
in which the shock front widths are of a few Debye lengths.

Even in a plasma without isolated non-neutral regions, the PIC or
kinetic simulations in general, have to resolve the Debye length to
suppress the inherent numerical instabilities when an explicit time
advancing scheme is
deployed~\cite{birdsall2018plasma,tajima2018computational}.  In a
fusion grade plasma with temperature $T_e\sim 10keV$ and density
$n_e\sim 10^{20}m^{-3}$, the Debye length is orders of magnitude
smaller as $\lambda_D/L\sim 10^{-6}$. Resolving the Debye length would
bring a substantial computational cost. This is especially so in a
plasma with strong density variation, for example, in a plasma TQ where the cold/dense edge plasma region has a far smaller Debye
length than the initial fusion-grade core
plasmas.~\cite{li2023staged}

One way to circumvent this obstacle is to perform downscaled PIC
simulations to decipher the physics scaling laws, in which the
characterized physical lengths including $\lambda_{mfp}$ and $L$ are
proportionally reduced compared to the Debye
length~\cite{li2023staged,Li-etal-prl-2022,li2022transport}. \textcolor{black}{This is to some extent equivalent to artificially enlarging the Debye length through the permittivity $\varepsilon_0$ in the Poisson equation while keep the other lengths unchanged~\cite{takizuka2017kinetic}.} A more
aggressive but related aim for downscaled simulations, which also
applies to experimental designs, is the concept of similarity.  For
example, the so-called ``perfect'' hydrodynamic similarity has been
argued by Ryutov and Remington~\cite{ryutov2003perfect} for simulating
astrophysical phenomena on laboratory laser facilities like the
National Ignition Facility. \textcolor{black}{Recent works on similarity and scaling
laws in the context of low-temperature plasma discharges can be found in Refs. \onlinecite{fu2021generalizing,duan2024evaluation,zheng2024scale,yang2025demonstration} and in the review paper Ref.~\onlinecite{fu2023similarity}}.

The current paper provides an analysis of our previous downscaled
kinetic simulation approach from the perspective of similarity.  We
will show that a simple similarity exists that provides a
straightforward scaling for the quasi-neutral plasmas but preserves
the non-neutral plasma structures. This appears to be a ``perfect''
solution in electrostatic plasma simulations to reconcile the system
size with the Debye length, while preserving the transport
physics. Specifically, under our scaling, the plasma distribution
function remains invariant almost everywhere, namely the
quasi-neutral region away from the isolated spots of non-neutral
plasmas. As a result, the density, velocity, pressure, and most
importantly, thermal conduction heat flux of plasmas (and neutrals if
present) are not scaled. In the fluid descriptions of plasmas, this
means that no approximation with regard to the equation of state is
needed for the similarity~\cite{ryutov2003perfect}.

Our downscaled simulation approach is different from those in
conventional similarity and scaling laws designed for experiments (see
Refs.~\onlinecite{ryutov2003perfect,buckingham1914physically,ryutov2018scaling,murakami2002scaling,falize2011similarity,fu2023similarity}
and references therein), in that the rates of collisions will be
scaled through an \textit{artificially} enhanced Coulomb Logarithm,
$\ln \Lambda$, in the collisional operators.  Rescaling the collision
rates only through the Coulomb algorithm allows the treatment of a wide
range of collisional processes including the three-body collisions in
plasma-neutral interactions.
Without losing generality, let's consider the electron Boltzmann
equation in a partially ionized plasma in the
electrostatic limit \textcolor{black}{without a background magnetic field $\mathbf{B}_0$ (the presence of $\mathbf{B}_0$ and the electromagnetic effect are discussed in the supplemental material), which applies to the transport in multi-dimensional
unmagnetized plasmas and the parallel transport in magnetized plasmas,}
 \begin{equation}
     \frac{\partial f_e}{\partial t} + \mathbf{v}\cdot\frac{\partial
       f_e}{\partial \mathbf{x}} + \frac{q_e}{m_e} \mathbf{E} \cdot
       \frac{\partial f_e}{\partial \mathbf{v}}=\sum_j
       C_{ej}(f_e,f_j,\textup{v}_{ej},\sigma_{ej}),\label{eq-Boltmann}
 \end{equation}
where $f_e(\mathbf{x},\mathbf{v},t)$ is the electron distribution and
$\mathbf{E}$ is the electric field.
Here $C_{ej}$ describes the collisions between electrons and
species $j$ (including neutrals) with all the collisional types. It
depends on the distributions $f_e$ and $f_j$, the relative
velocity $\textup{v}_{ej}$, and the cross-section of the process
$\sigma_{ej}$. Similar Boltzmann equations can be obtained for other
species' distribution functions $f_j$.

One interesting and important property of the collisional operators
$C_{ej}$ is that they are proportional to the collisional
rates~\cite{krasheninnikov2020edge} defined as
$\mu_{ej}n_e=\int_0^\infty
\sigma_{ej}(v)v_{ej}f_e(\mathbf{v})d\mathbf{v}\propto \ln
\Lambda$. Such rates appear naturally since the collisional terms on
the right-hand-side (RHS) of the Boltzmann equation represent the
rates of the change of distribution function, $\delta f_e/\delta t$,
in a given region of phase space as a result of collisions.

Under the transformation of
\begin{align}
  t \rightarrow t/\mathscr{L},\,\,\,
  \textbf{x}
  \rightarrow  \textbf{x}/\mathscr{L}, \,\,\,
  \mathbf{E}\rightarrow
  \mathscr{L}\mathbf{E}, \,\,\,
  \mu \rightarrow \mathscr{L}\mu, \label{eq:ideal-scaling}
\end{align}
the Boltzmann equation in the electrostatic limit,
Eq.~(\ref{eq-Boltmann}), preserves the solution of the distribution
function $f_e$ and $f_{j\neq e}.$ This can be easily seen by
multiplying $\mathscr{L}$ and reformulating the Boltzmann
equation. \textcolor{black}{The scaling of the collisional rate $\mathscr{L}\mu$ in Eq.~(\ref{eq:ideal-scaling}) can be realized via scaling the Coulomb Logarithm $\mathscr{L}\ln \Lambda$ in the plasma simulations}, so relative strengths of different
collisional processes are preserved.  By downscaling the system size with
the same factor while holding the plasma speed unchanged, the ratio of
the collisional mean free path and the system size or the Knudsen
number is unchanged. In cases where the Knudsen number is the critical
dimensionless parameter to set the transport physics, for example, in the
TQ problem~\cite{li2023staged}, one can intuitively
anticipate the transport physics will be fully captured in the
downscaled simulations.

Since the distribution function and particle velocity
are both invariant, the moments of the distribution function are
invariances as well. These include the density, velocity, temperature,
and most interestingly, the electron conduction heat flux, the proper
closure of which is required in the fluid modelings of
plasmas~\cite{braginskii,Bell-pof-1985,zhang2023cooling}. In the
collisional limit, the Braginskii equations for nearly Maxwellian
plasmas~\cite{braginskii} have both the conduction heat flux
$\mathbf{q}$ and stress tensors $\Pi$ invariant under the
transformation of Eq.~(\ref{eq:ideal-scaling}).  Similarly, the
electron conduction flux $\mathbf{q}_e$ in the collisionless regime is
also an invariance, either in the free-streaming
limit~\cite{Bell-pof-1985}, $q_e\propto n_eT_{e\parallel}v_{th,e}$
with $v_{th,e}$ the electron thermal speed, or in the convective
scaling regime~\cite{zhang2023cooling}, $q_e\propto
n_eT_{e\parallel}V_{i\parallel}$ with $V_{i\parallel}$ the ion flow
speed.

Another interesting and important property of the similarity enabled
by the transformation Eq.~(\ref{eq:ideal-scaling}) of the
electrostatic Boltzmann equation~(\ref{eq-Boltmann}) is that the
collisional processes, except for the collisional rates,
 are invariant due to the invariances of the distribution functions.
 As a result, the downscaled simulations can be applied to a variety
 of collisional plasmas, e.g., the partially ionized plasma where the
 nonlinear three-body-type collisions are
 important~\cite{krasheninnikov2020edge}. Because the scaled-up
 collisional rates include the ionization and excitation rates, the plasma
 cooling rates due to both the ionization and the radiation scale up accordingly, while the
 ``ionization cost,'' or the energy dissipation per ionization
 event~\cite{krasheninnikov2020edge}, is an atomic physics quantity
 and unscaled.  
 As
 a result, the plasma energy density change remains the same due to
 the downscaled time, though the total plasma energy and its change are
 scaled down with the system size.

In practice, our downscaled simulation only has explicit control over
the system size and the collisional mean free paths, both
of which rescale with the same factor $\mathscr{L}.$ The initial particle
distribution $f(\mathbf{v})$ is not modified.  The rescaling of time
and electric field are the outcome of the downscaled
simulation. Since both the particle transit time $L/v$ and the
collisional time are scaled down by the same factor $\mathscr{L},$ the
dynamical time in the simulation would be scaled down by
$\mathscr{L},$ if the associated transport physics is set by the Knudsen number
alone. The behavior of the electric field is the most intriguing.
In the bulk of the plasma simulation domain, the plasma is quasi-neutral, in which the ambipolar electric field is approximated from the electron momentum equation, \textcolor{black}{e.g.,
by balancing the electric field force with the electron pressure gradient and the thermal force for nearly-equilibrium plasmas~\cite{braginskii}
\begin{align}
e\mathbf{E} \approx - \frac{1}{n_e} \frac{\partial (n_e T_e)}{\partial\mathbf{x}}-0.71n_e\nabla_\parallel T_e.
\end{align}}
Since the density and temperature are invariances, this produces the
rescaling $\mathbf{E} \rightarrow \mathscr{L}\mathbf{E} ,$ as desired in Eq.~(\ref{eq:ideal-scaling})
for a ``perfect'' similarity.

The situation in isolated non-neutral regions is quite different. Since
both $n_e$ and $T_e$ are invariant under the ideal transformation of
Eq.~(\ref{eq:ideal-scaling}), the Debye length $\lambda_D$ is
unchanged.  As a result, the electric field in the non-neutral
region, which has the original physics scaling of $T_e/\lambda_D,$
is unaffected in our downscaled simulation. This is actually a desired
outcome in that non-neutral plasma region, such as the plasma sheath or
the double layer structure in a plasma shock, is tiny in space $\sim
\lambda_D$ compared with the system size $L.$ Such an extreme
scale separation $L \gg \lambda_D$ presents a natural micro-macro
decomposition of the physics.  Our downscaled simulations now keep the
microscopic physics at its original scale, but preserve the macroscopic
transport physics through a similarity transformation.

The scale-up of the macroscopic or quasi-neutral electric field
$\mathbf{E}$ by a large factor of $\mathscr{L},$ as required for the
similarity between the full-sized system and the downscaled
simulation, comes at the expense of a scaled-up charge separation in a
quasi-neutral plasma.  This can be seen from the Poisson's equation,
\begin{equation}
    \nabla \cdot \mathbf{E} =\frac{q_e}{\epsilon_0} \int (f_e-f_i) d\mathbf{v}. \label{eq-gaussian-equation}
\end{equation}
We should highlight again that the electric field scaling can not be obtained from the Poisson's equation~\cite{fu2023similarity}, but an automatic outcome of the ambipolar transport.  Expressed in terms of Debye length, one finds that ambipolar potential
variation $\Delta \Phi$ in the quasi-neutral plasma, which has a
gradient length scale of $L,$ takes the physical scaling,
\begin{align}
\frac{e \Delta \Phi}{k_B T_e} \sim \frac{L^2}{\lambda_D^2} \frac{\delta n}{n_0}.
\end{align}
Here the plasma potential has the physical scaling $\Delta\Phi \sim
T_e,$ so it is an invariant under the transformation of
Eq.~(\ref{eq:ideal-scaling}).  In a real plasma, the charge separation
($\delta n = n_i-n_e$ for a hydrogen plasma) is usually tiny to set up
the quasi-neutral electric field,
\begin{align}
  \frac{\delta n}{n_0} \sim \frac{e \Delta\Phi}{k_B T_e} \frac{\lambda_D^2}{L^2} \sim \frac{\lambda_D^2}{L^2},
\end{align}
for the smallness of $\lambda_D/L.$ In a downscaled simulation, $L$
is shrunk by a factor of $\mathscr{L}$ so the effective charge
separation is artificially enhanced by a factor of $\mathscr{L}^2.$

It may be argued that such enhancement of $\delta n/n_0$ may be
desirable in down-scaled PIC simulations. This is because the PIC code
employs the so-called macro-particles that usually represent
aggregates of real particles to meet the memory and computation
efficiency constraints~\cite{birdsall2018plasma}. As such, the minimum
amount of numerical charge separation in the PIC simulations is
limited as \textcolor{black}{$(\delta n/n_0)_{\rm min}\approx 1/N$} with $N$ being the number
of macro-particles in the cell, which must be larger than \textcolor{black}{$(\delta
n/n_0)_{\rm min}$} in reality. So an enhancement of a physical $\delta n$ can
reduce the inaccuracy induced by the limited amount of numerical
macro-particles.  This potential saving in the number of
macro-particles required by accuracy is in addition to the
computational cost reduction by a shrunk simulation domain compared
with $\lambda_D$ and a shortened dynamical time scale or collisional
time $\tau_c$ compared with the inverse of the plasma frequency
$\omega_{pe}^{-1}.$ Using a one-dimensional configuration space as an
example, the total simulation savings simply due to this rescaling of
space and time would be a factor of $\mathscr{L}^2.$

The weakly collisional or nearly collisionless plasmas allow
long-living plasma waves in the electrostatic limit, so the plasma
wave-particle interactions can involve additional characteristic
lengths and time scales associated with a family of waves with
distinct frequencies and wavelengths~\cite{stix1992waves}. One
immediate consideration of a downscaled simulation for such a plasma
is that the wave frequency, including the real frequency $\omega_r$
and the growth/damping rate $\gamma$ with $\omega=\omega_r+i\gamma,$
and wavelength ($\lambda=2\pi/k$) are not straightforwardly scaled
depending on the wave/instability types. Specifically, since the
configuration space is scaled but not the velocity space, the
instabilities driven by gradients in configuration space like the
Rayleigh-Taylor and resistive drift wave
instabilities~\cite{zhang2020different} should be scaled with
$\mathscr{L}\omega$ and $\mathscr{L}k$ (note that for the
Rayleigh-Taylor instability, the effective gravity, $g$, should be
scaled as $\mathscr{L}g$). Whereas, the velocity space instabilities
like beam
instabilities~\cite{stix1992waves,dawson1960plasma,buneman1963excitation}
are not scaled since the distribution function and hence the
instability drive is an invariance under the transformation of
Eq.~(\ref{eq:ideal-scaling}). As a result, the scaled transport and
collisional time may affect the wave properties that can in turn alter
the plasma dynamics, e.g., a largely enhanced collisional rate can
cause a significant increase of the collisional damping rate of the
velocity space instabilities~\cite{auerbach1977collisional}. It should
be noted that even though the wave frequencies and wavelengths can be
scaled for the configuration space instabilities, the plasma-wave interactions still remain uncertain due to the
nonlinear saturation of the instabilities. All these effects will be
manifested in the downscaled simulations with different $\mathscr{L}$, so the subtlety
they can potentially bring to the transport physics can be straightforwardly checked
in a $\mathscr{L}$ scan of downscaled simulations.

Lastly, to ensure that the scaled systems are identical to the
original one, all initial and boundary conditions must be scaled
consistently.  While the initial condition can be straightforwardly
scaled, e.g., $f(\mathbf{x}/\mathscr{L},\mathbf{v})$ for the
distribution functions, the boundary conditions may differ depending
on the specific physical problems. Although it is impractical to list
all possible boundary conditions for plasma applications, commonly
used ones, such as the periodic~\cite{zhang2023collisional}, absorbing~\cite{Li-etal-prl-2022},
and various kinds of recycling~\cite{li2023staged,li2023plasma}
boundary conditions, generally do not require scaling. For example,
the absorbing and recycling boundary conditions depend on the particle
flux to the simulation boundaries, which is invariant in this
similarity. Although the particle numbers through the boundaries are
scaled down due to the reduction of the simulation time, their
fractions to the total particle numbers are invariant.

As a demonstration of the proposed similarity, we will consider two
one-dimensional (1D) plasma problems using VPIC code~\cite{VPIC}.
These are dynamic plasma TQ~\cite{li2023staged} and the steady-state
plasma sheath in a high-recycling divertor~\cite{li2023plasma}, which are ideal for demonstrating the similarity
properties. Particularly, the plasma in the TQ undergoes a dynamic
cooling from the collisionless to the collisional regimes and thus
different closures of the electron thermal conduction will be
included~\cite{li2023staged}. It also involves a locally propagating
non-neutral shock structure in the collisionless
stage~\cite{zhang2023cooling}. In the plasma sheath problem,
the separation of the non-neutral Debye sheath and the quasi-neutral
presheath is evident and the Bohm criterion applies
to a spatially extended transition layer~\cite{Li-etal-prl-2022}. More importantly,
the plasma sheath in a high-recycling divertor includes complex
plasma-neutral interactions such as elastic collision, excitation,
ionization, and charge-exchange~\cite{li2023plasma}. In addition, both
of these examples contain nontrivial plasma recycling boundary
conditions with different types. \textcolor{black}{The VPIC code and simulation setups can be found from Refs.~\onlinecite{li2023staged,li2023plasma,VPIC}, which are summarized in the supplemental materials.}

\begin{figure}[!htb]
    \centering  
    \subfloat[]{\includegraphics[width=0.8\linewidth]{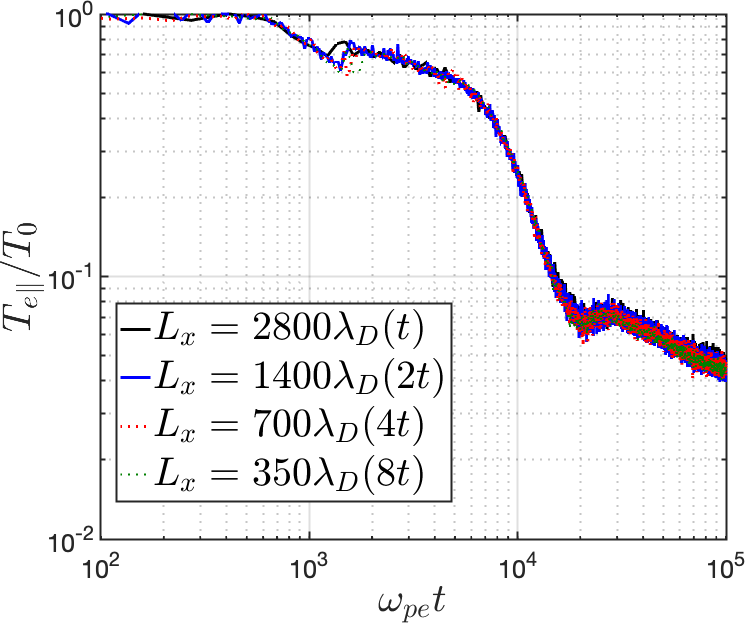}}\\
    \subfloat[]{\includegraphics[width=0.8\linewidth]{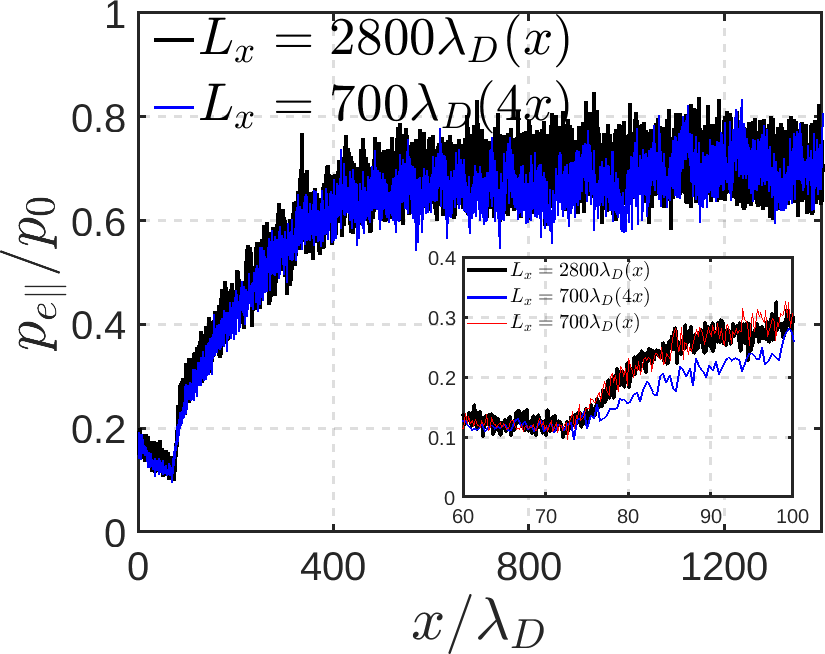}}
    \caption{Results of four scaled TQ simulations with fixing $K_n=\lambda_{mfp}/L_x =98$. (a) Time evolution of the center parallel electron temperature $T_{e\parallel}$, in which the time for small $L_x$ cases ($L_x<2800\lambda_D$) have been rescaled accordingly. (b) The parallel electron pressure profiles at $\omega_{pe}t=2000$ for $L_x=2800\lambda_D$ (thick black) and at $\omega_{pe}t=500$ for $L_x=700\lambda_D$ (thin blue), where the spatial coordinate for the latter is quadrupled. In the zoom-in figure near the shock front around $x=74\lambda_D$, we also plot $p_{e\parallel}$ for $L_x=700\lambda_D$ case (thinnest red) with the original spatial coordinate (but shifted so that its shock front co-locates with the other two). A reduced ion-to-electron mass ratio of $m_i=100m_e$ is employed. The Debye length for the initial hot plasma,
$\lambda_D\propto\sqrt{T_0/n_0}$ is used for normalization. }
    \label{fig-thermal-quench}
\end{figure}

\textcolor{black}{In the plasma TQ problem, the initial 1D nearly-collisionless plasma has a fixed Knudsen number of
$K_n=\lambda_{mfp}/L_x=98\gg 1$. The cooling boundary will turn the plasma temperature down by 100 times so that the plasma is initially nearly-collisionless but eventually collisional as a result of cooling.} The results of four similar simulations
with different $L_x$ are shown in Fig.~\ref{fig-thermal-quench},
demonstrating nearly identical plasma TQ processes. Particularly, the
time evolution of the center electron temperature in
Fig.~\ref{fig-thermal-quench}(a) indicates that the electron thermal
conduction flux should be an invariance in both the collisionless
($\omega_{pe}t<2\times 10^4$) and collisional regimes
($\omega_{pe}t>2\times 10^4$). While the electron pressure
$p_{e\parallel}=n_eT_{e\parallel}$ profiles in
Fig.~\ref{fig-thermal-quench}(b) provide another straightforward
evidence of the similarity. As aforementioned, $p_{e\parallel}$ also
manifests the electric field under the ambipolar transport constraint in the nearly-collisionless plasma
since its gradient balances with the electric
force~\cite{zhang2023cooling}, i.e., $E_x\propto \partial
p_{e\parallel}/\partial x$. In fact, it is more practical to use
$p_{e\parallel}$ rather than $E_x$ itself to illustrate the scaling
laws of the electric field since $E_x$ suffers Langmuir perturbations
as shown in Fig.~\ref{fig-thermal-quench}(b) that is not invariant. We
see that $p_{e\parallel}$ in the bulk plasma is nearly the same, reminiscent of an enhancement of $E_x$ in the cases with small $L_x$. More importantly, the zoom-in plot near the shock front
around $x=74\lambda_D$ shows that the shock front structure and hence
the associated electric field is not scaled by comparing the
$p_{e\parallel}$ profiles in the original coordinates (the thick black
and thin red curves).

\begin{figure}[!htb]
    \centering
    \includegraphics[width=0.9\linewidth]{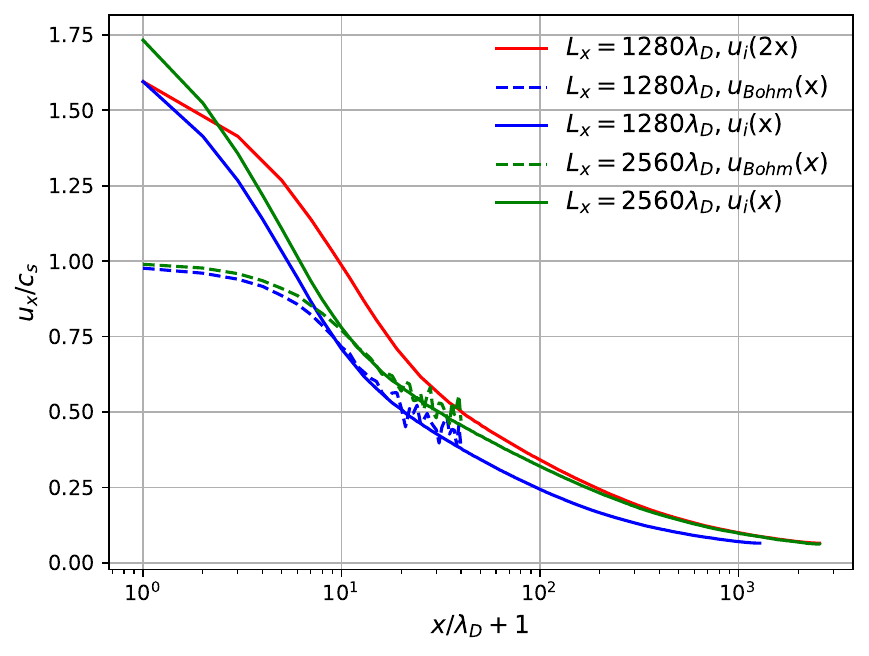}
    \caption{Ion exit flow and the Bohm speed in Eq.~(\ref{eq:ubohm}), normalized by local $c_s=\sqrt{(T_{ex}+3T_{ix})/m_i}$, in two scaled sheath simulations. For the $L_x=1280\lambda_D$ case, both the original and rescaled results in terms of the spatial coordinates are included. Here $\lambda_D$ is for a plasma with $T_e=10eV$ and $n_e=5\times 10^{19}m^{-3}$. }
    \label{fig-sheath}
\end{figure}

In the plasma sheath simulations, a
power input, $Q_{in}$, at the right boundary is utilized to supply the power for the plasma cooling and neutral ionization in the whole domain. The energy input should be scaled with the total particles and hence the system domain, but not the power input due
to the scaled time. Thus, this power input through a unit surface area from the boundary is different from the power density that should be scaled up just as the radiative cooling rate. Here $Q_{in} = 10 MW/m^2$ is used for an averaged
plasma density of $\overline n = \int_0^{L_x} (n_i+n_n) dx/L_x=
10^{20} m^{-3}$ with $n_i$ and $n_n$ being the ion and neutral
density, respectively. In such a problem, a complex Bohm criterion that
gives the lower bound of the ion exit flow speed of $u_{Bohm}$ apples
to a wide transition layer by considering all the transport
physics~\cite{li2023plasma}
\begin{equation}
    u_{ix} \geq u_{Bohm} \equiv \sqrt{(3T_{ix}+\beta_{Bohm} T_{ex})/m_i},
    \label{eq:ubohm}
\end{equation}
where $\beta_{Bohm}$
\begin{equation}
   \beta_{Bohm} \equiv -\frac{3\left(\cfrac{\partial q_n^i/\partial \phi}{en_{i}u_{ix}} -1 + \cfrac{J_{ixx} }{e\Gamma_{i}E}-\cfrac{I_{ix}}{en_iE} \right)}{\cfrac{\partial{q_n^e}/\partial\phi}{en_{e}u_{ex}} + 1 + \cfrac{J_{exx} }{en_{e}u_{ex}E}-\cfrac{I_{ex}}{en_eE} 
   }.
   \label{eq:beta}
\end{equation}
Here $q_n$ is the thermal conduction flux of $x-$degree of freedom in
the $x$ direction, and $I_x$ and $J_{xx}$ is, respectively, the sum of
momentum and energy change in the $x$ direction due to all collisions. All
these terms can be evaluated locally in VPIC simulations, where
time-averaged plasma parameters over a long time period in the steady
state are utilized to suppress the PIC noises in contrast to the post-process denoising approach~\cite{picklo2024denoising}. Two simulations
with different $L_x$ and corresponding mean free paths are conducted,
with the steady-state ion exit flow and Bohm speed $u_{Bohm}$ shown in
Fig.~\ref{fig-sheath}. It shows that the plasma behaviors in the bulk
region $x\gtrsim 100\lambda_D$ are nearly identical by comparing the
rescaled results (the red and the green). While the plasma sheath is
almost not scaled (e.g., see the blue and the green), the nominal
entrance of which is denoted by the separation of $u_i$ from
$u_{Bohm}$.

In conclusion, a simple but powerful similarity was discussed for
kinetic simulations of electrostatic plasmas, in which the
characteristic lengths including the system size and mean free paths
are reduced in contrast to the small invariant Debye length. The
similarity properties and its applicability have been discussed. The
utility of this similarity has been showcased by VPIC simulations of
the dynamic TQ and the steady-state plasma (pre)sheath in a high
recycling divertor.

\textbf{Acknowledgment} We thank the U.S. Department of Energy Office
of Fusion Energy Sciences through the Base Fusion Theory Program at Los Alamos National Laboratory (LANL)
under contract No. 89233218CNA000001. This research used resources of the National Energy Research Scientific Computing Center, a DOE Office of Science User Facility
supported by the Office of Science of the U.S. Department of Energy
under Contract No. DE-AC02-05CH11231 using NERSC award
FES-ERCAP0032298 and LANL Institutional Computing Program, which is supported by the U.S. Department of Energy National Nuclear Security Administration under Contract No. 89233218CNA000001.

\bibliography{reference}

\section*{Supplemental material to ``Similarity for downscaled kinetic simulations of electrostatic plasmas: reconciling the large system size with small Debye length"}

\section{VPIC code, simulation setups and diagnostics}
VPIC~\cite{VPIC} is a general purpose particle-in-cell simulation code for modeling kinetic plasmas in one, two, or three spatial dimensions. It employs a second-order, explicit, leapfrog algorithm to update charged particle positions and velocities in order to solve the relativistic kinetic equation for each species in the plasma, along with a full Maxwell description for the electric and magnetic fields evolved via a second-order finite-difference-time-domain (FDTD) solve. 
A Monte Carlo collision (MCC) module with null collision method~\cite{nanbu2000probability} is developed in VPIC to handle the collisions~\cite{li2023plasma}. In such module, a constant Coulomb Logarithm is provided by the user to quantify the collisional rates.

In the plasma TQ problem~\cite{li2023staged}, an initial 1D uniform plasma with
temperature $T_0= 10keV$ and density $n_0=10^{20}m^{-3}$ is bounded by
the thermobath boundaries at the two ends~, which recycle plasma
particles but clamp the temperature of the recycled plasma particles
to $T_w=0.01T_0$. The
cell size and number of macro-particles per cell are the same in
different simulations, except for the smallest $L_x=350\lambda_D$ case
where doubled macro-particles per cell have been used to reduce the
simulation noise, which reaffirms the advantage of system size
reduction. The initial Knudsen number is fixed as
$K_n=\lambda_{mfp}/L_x=98\gg 1$ so that the plasma is initially nearly
collisionless but eventually collisional as a result of cooling
towards the temperature $T_w.$

For simulating the plasma sheath in a high-recycling divertor, a
``closed box'' setup is employed following
Ref.~\onlinecite{li2023plasma}: a recycling tungsten wall is
implemented at the left boundary of the simulation domain with a unit
total particle recycling coefficient (including the reflected
particles associated with a comprehensive wall energy recycling coefficient and desorbed particles with wall temperature); while a
power input at the right boundary is utilized to supply the power for
the plasma cooling and neutral ionization in the whole domain. Interested readers are recommended to read Ref.~\onlinecite{li2023plasma} for detailed implementation.

In the simulations, the plasma density, parallel flow, parallel temperature and thermal conduction heat flux of the parallel degrees are calculated following the statistic physics from the particle distribution function $f$,
\begin{align}
 n=&\int f d^3v, \\
nV_x=&\int v_x f d^3v, \\
 nT_x=& m\int (v_x-V_x)^2f d^3v, \\
 q_{n}=&m\int (v_x-V_x)^3 f d^3v.
\end{align}
 Thus, the ``temperature'' cannot be
  used to describe the ``slope'' of the total electron
  distribution but measures the kinetic energy spread around a mean flow for
  particles.

\section{Background magnetic field and electromagnetic effects}
The downscaled electrostatic kinetic simulations can offer remarkable property of near-perfect similarity in multi-dimensional unmagnetized plasmas or in 1D magnetized plasmas along the magnetic field, as shown in the
main body of the paper. In the electrostatic simulations of multi-dimensional magnetized plasmas, the PIC approach may need to be modified though the similarity still holds. To see this formally, we recall that the
Boltzmann equation with magnetic field has the form
\begin{align}
  \frac{\partial f_e}{\partial t} + \mathbf{v}\cdot\frac{\partial
    f_e}{\partial \mathbf{x}} + \frac{q_e}{m_e} \left(\mathbf{E} + \mathbf{v}\times\mathbf{B}\right) \cdot
  \frac{\partial f_e}{\partial \mathbf{v}}=\sum_j
  C_{ej}(f_e,f_j,\textup{v}_{ej},\sigma_{ej}),\label{eq-Boltmann-Maxwell}
\end{align}
where $\mathbf{B}$ is a constant in the electrostatic limit but can change in time to account for the electromagnetic effects. For $f$ to remain unchanged under the transformation, the magnetic field must follow the
transformation or rescaling of $\mathbf{B} \rightarrow
\mathscr{L}\mathbf{B}.$ With a background magnetic field, this will alter several plasma properties
including the plasma beta, $\beta\propto n_eT_e/B^2\propto1/\mathscr{L}^2$, the Alfv\'en
speed, $V_A\propto B/\sqrt{n}\propto\mathscr{L}$, and the particle gyro-radii $\rho_{e,i}\propto v_{th,e,i}/\omega_{ce,i}\propto 1/\mathscr{L}$, etc.

In the electrostatic limit, these modified quantities by $\mathbf{B}$ may not affect the transport physics. For example, the particle drift velocities are not scaled but the banana orbit width is reduced by a factor of $\mathscr{L}$ in tokamaks so that the neoclassical transport is invariant~\cite{helander2005collisional}. 

However, there are two potential issues in the multi-dimensional magnetized plasmas even in the electrostatic limit, which can appear in the electromagnetic plasmas as well though the electromagnetic effects are more complicated. The first issue is that the multi-dimensionality of the plasma may induce new instabilities, like the resistive drift wave instabilities in tokamaks~\cite{zhang2020different}, that introduce new time and length scales into the system. These instabilities may depend on the background magnetic field or not. As discussed in the main body of the paper, the instabilities and associated waves themselves already challenge the applicability of the proposed similarity, while the scaling of $\mathscr{L}\mathbf{B}$ simply make the situation more complicated. The second issue arises when the plasma has a large electron gyrofrequency, $\omega_{ce}\gtrsim \omega_{pe}$. In such case, the simulation time step, $\Delta t$, in the scaled simulations with much enhanced $\mathbf{B}$ and hence $\omega_{ce}$ is limited by the requirement of resolving the gyro-motion ($\omega_{ce}\Delta t<1$) rather than the CFL constraint. Thus, the saving factor is reduced, e.g., from $\mathscr{L}^2$ to $\mathscr{L}$ in 1D case. One possible solution to this issue can be applying the guiding-center
motion for electrons to avoid the requirement of resolving $\omega_{ce}$~\cite{takizuka2017kinetic}.

In the electromagnetic limit, the application of this scaling is more subtle. For cases where the physics is not dependent on the aforementioned scaled variables related to $\mathbf{B}$, the scaling of $\mathscr{L}\mathbf{B}$ still works, as in Ref.~\cite{wang2023similarity}. However, for the plasma transport that indeed depend on the plasma beta, Alfv\'en speed, etc., the downscaled kinetic simulations can no longer capture the correct physics. For such case, one can alternatively decide not to scale
$\mathbf{B}$ so the plasma beta stays the same in the downscaled
simulations. Whether the relevant physics is properly captured in
such simulations requires case-by-case examinations. An illustrative example we have encountered is again in the plasma
thermal quench problem.  The issue is that in the electrostatic limit,
plasma cooling as a result of rapid parallel transport to the
boundary, as envisioned for a globally stochastic magnetic field,
brings down the $T_{e\parallel}$ rapidly but not the $T_{e\perp}$ (e.g., see Fig.~\ref{fig-Tepara-Teperp} in the supplemental materials). An
intriguing question is whether there are collisionless cooling
mechanisms that can bring down $T_{e\perp}$ at the rate by which
$T_{e\parallel}$ is cooled by the parallel
transport~\cite{zhang2024collisionless}.  One plausible candidate is
the velocity-space whistler wave instability, previously known from
Ref.~\cite{guo2012ambipolar}, for a truncated electron distribution,
which is an electromagnetic wave instability.  Although the strength
of the magnetic field does not affect the formation of the truncated
electron distribution and hence $T_{e\parallel}$ cooling, which is due to tail electron loss through
parallel transport, it does impact the dispersion of the unstable
whistler modes,~\cite{kennel1966limit} so the background magnetic
field should not be scaled.  For such a velocity-space instability,
rescaling the system with the exception of the background magnetic
field thus has no effect on the dispersion and growth rate of the
modes, as long as the wavelength is short compared with the downscaled
system size.  Again, the benefit of such downscaled electromagnetic
simulations without B-rescaling is particularly great for this problem since $\omega_{ce}\gtrsim \omega_{pe}$. As a result,
the perturbed quantities including the electromagnetic fields and
current are invariances, although one needs to care about the enhanced
collisional damping effect in the downscaled
system~\cite{zhang2024collisionless,zhang2023collisional}.

\begin{figure}[!htb]
    \centering
\includegraphics[width=0.8\linewidth]{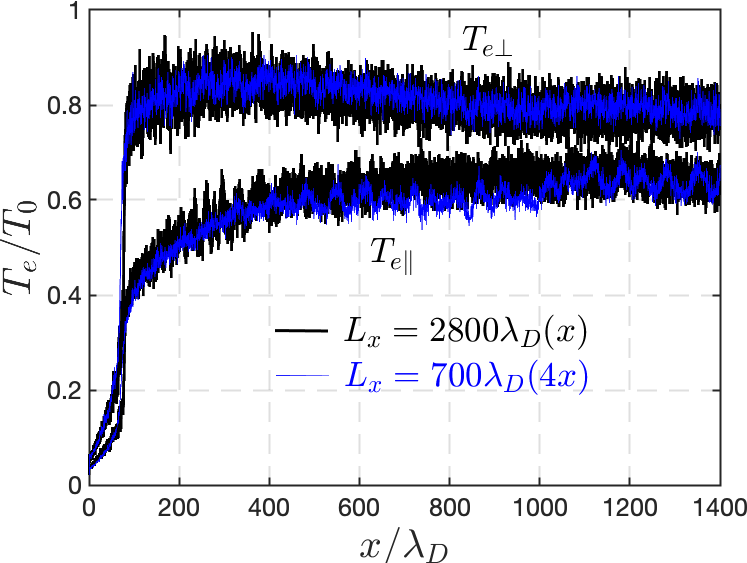}
    \caption{The parallel and perpendicular electron temperature in the thermal quench simulations, corresponding to Fig.~1b in the main body of the paper. }
    \label{fig-Tepara-Teperp}
\end{figure}

\end{document}